\documentclass{article}
\usepackage{latexsym}
\usepackage{graphicx}

\begin{document}

% You may use Title,Subject,Author,Manager,Company,Operator,
% Category,Comment,Hlinkbase document properties here
\title{Derivation of the General Case Sagnac\\Result using Non-time-orthogonal Analysis}
\author{Robert D. Klauber\\1100 University Manor Dr. 38B, Fairfield, IA 52556\\rklauber@.netscape.net}
\date{Final revision acceptance June 11, 2003}
\maketitle

\begin{abstract}

The Sagnac time delay and fringe shift dependency on angular velocity and 
enclosed area are derived from the rotating reference frame using 
non-time-orthogonal (NTO) tensor analysis. NTO analysis differs from 
traditional approaches by postulating that the continuous and single valued 
nature of physical time constrains simultaneity in a rotating frame to be 
unique (and thus not a matter of convention.) This implies anisotropy in the 
physical, local speed of light and invalidity of the hypothesis of locality 
for NTO frames. The Sagnac relationship for the most general case, in which 
the area enclosed is not circular and does not have the axis of rotation 
passing through its center, is determined.

\bigskip

\end{abstract}

\bigskip

\section{INTRODUCTION}
\label{sec:introduction}
In 1913 M. G. Sagnac\cite{Sagnac:1914} sent light rays in the clockwise 
(cw) and counterclockwise (ccw) directions on a rotating platform, and 
examined the resulting interference fringe patterns made by the returning 
rays. He found, as have subsequent researchers, that the fringes shifted as 
the angular velocity of the platform changed. Post\cite{Post:1967} 
summarized the results of these researchers and presented the experimentally 
determined first order relationship
\begin{equation}
\label{eq1}
\Delta Z=\frac{\Delta \lambda }{\lambda _0 }=\frac{4\mbox{\boldmath$\omega$} \cdot \mbox{\boldmath$A$}}{c\lambda 
_0 },
\end{equation}
where \textbf{$\omega $} and \textbf{A} are angular velocity and area vector 
quantities, respectively, $c$ is the standard value for speed of light in 
vacuum/air, and $\Delta Z$ is the fractional change in fringe location. 
$\lambda _{o}$ is wavelength of the light in the lab frame before it is 
directed down the axis of rotation, out a radial direction on the rotating 
platform, and split via a 50{\%} reflection/transmission mirror into the two 
cw and ccw rays. In the simplest case where the two rays travel circular 
paths $A = $2$\pi r$ and $\omega =v/r$.

From the standpoint of special relativity theory (SRT), this result may seem 
perplexing, as SRT posits that no change in speed of the reference frame 
should affect the speed of light or any measurements that one could make. 
This principle is considered, via the hypothesis of locality (sometimes 
``the surrogate frames posulate''), to extend to general relativity (GR) in 
which (for the Einstein synchronization of SRT) the \textit{local physically measured} one-way speed of light 
is $c$, even in non-inertial frames. 

The hypothesis of locality, a linchpin in the traditional approach to 
relativistic rotation, states that a local inertial observer is equivalent 
to a local co-moving non-inertial observer in matters having to do with 
measurements of distance and time. (See M{\o}ller\cite{ller:1969}, 
Einstein\cite{Stachel:1950}, and Mashoon\cite{Mashoon:2003}.) It follows 
immediately that a Lorentz frame can be used as a local surrogate for the 
non-inertial frame, and in such a frame (with Einstein synchronization), the 
local, measured, one-way speed of light is $c$.

Minguzzi\cite{Minguzzi:2003} and M{\o}ller\cite{See:1}, among others, note 
that the hypothesis of locality is only an assumption. It is, however, an 
assumption that, historically, has worked very well in a large number of 
applications.

Given the foregoing, experiments performed in non-inertial frames, like 
those in inertial frames, should not have results dependent on the speed of 
the experimenter. Yet in the Sagnac experiment, the result depends on the 
speed $v$ of the platform along the path traveled by the rays.

Consider, for example, that the cw and ccw light rays travel identical 
routes (in reverse), and according to SRT (and GR) each must have the same 
speed $c$ as physically measured locally at all points along that path 
regardless of the motion of the platform. Hence, it seems one could only 
conclude that wave maxima (or minima) on both rays should return to the 
fringe location at the same time, regardless of path speed $v$, in apparent 
disagreement with the Sagnac result.

One may at first consider that the light rays originated from one source in 
the lab and that as they struck the half silvered mirrors and were reflected 
in opposite directions, they may have undergone Doppler shifting, which 
resulted in the observed effect\cite{Dresden:1979}. However, Dufour and 
Prunier\cite{Dufour:1942} carried out a series of experiments with the 
light source located on the platform itself and found no change in observed 
results.

Furthermore, the thought experiment of Appendix A demonstrates that short 
light pulses emitted from the same point on the platform, traveling the same 
path in the cw and ccw directions, must necessarily arrive back at that 
point at different times. Thus, any observer at that point would seem 
constrained to the conclusion that light, as seen from a rotating frame, 
travels at different speeds in the cw and ccw directions.

From the lab frame, derivation of the arrival time difference between the cw 
and ccw light pulses is well 
known\cite{Selleri:1997}$^{,}$\cite{Rizzi:1998}$^{,}$\cite{Bergia:1998} 
(summarized, for reference, in Appendix B), and from it, (\ref{eq1}) readily results 
(see Section \ref{subsec:circular}.) From the rotating frame, 
however, the situation is less clear. If one insists on the speed of light 
being locally invariant and isotropic, then it is not readily deduced that 
the cw and ccw rays arrive back at the emission point at different times.

Selleri\cite{Selleri:1997} has made this derivation from the rotating 
frame, but only by denying the invariance of one way light speed in 
\textit{inertial }frames, and thus effectively dismantling relativity theory as we know it. 
Others\cite{Rizzi:1998}$^{,}$\cite{Bergia:1998}$^{,}$\cite{Rizzi:1999}$^{,}$\cite{Rizzi:1} 
have based a derivation on local invariant/isotropic light speed, and a 
``desynchronization of clocks''. A satisfactory description of that approach 
is beyond the scope of this article, though the author suggests it is 
inappropriate physically. In his opinion, it leads to a discontinuity in 
physical time, clocks out of synchronization with themselves, and an 
arbitrary number of possible time readings on the same standard clock in the 
same coordinate frame for the same event. (See 
Klauber\cite{Klauber:2003}$^{,}$\cite{Klauber:1998}.)

The present author has carried out an analysis of rotating 
frames\cite{Klauber:2003}$^{,}$\cite{Klauber:1998}, similar in some 
aspects to an earlier and less extensive analysis by 
Langevin\cite{Langevin:1937}, using the metric obtained from the most 
commonly accepted transformation between the lab and the rotating frame. 
This metric has off diagonal time-space components and implies a 
non-orthogonality between time and space in rotating frames, dubbed by the 
author ``non-time-orthogonality''.

That metric has been used successfully in the global positioning system 
(GPS) to account for time delays that would not be predicted under the usual 
relativistic assumption of isotropic local light speed. In fact Neil Ashby, 
recognized leader in GPS analysis, notes \textit{`` .. the principle of the constancy of c cannot be applied in a rotating reference frame ..''}$^{ }$\cite{Ashby:2002}. He also 
states \textit{``Now consider a process in which observers in the rotating frame attempt to use Einstein synchronization (constancy of the speed of light) ..... Simple minded use of Einstein synchronization in the rotating frame ... thus leads to a significant error''}.\cite{Ashby:1997} 

In non-time-orthogonal (NTO) analysis, one does not insist on transforming 
to local time orthogonal frames on the rotating platform, as is common among 
prior researchers. Instead, one simply examines the NTO metric and from it, 
deduces concomitant physical world behavior. When this is done, not only can 
the Sagnac effect be derived (as shown below), but all other observed 
rotating frame effects can be, as well. (See 
Klauber\cite{Klauber:2003}$^{,}$\cite{Klauber:1998}$^{,}$\cite{Klauber:1999}$^{,}$\cite{Klauber:1}$^{,}$\cite{Klauber:2}$^{,}$\cite{Klauber:3}$^{,}$\cite{Klauber:4}$^{,}$\cite{Klauber:5}.) 
Further, the geometric foundation of relativity theory remains intact. All 
general and special relativity analyses for time orthogonal frames are 
unchanged. That is, NTO analysis reduces to the traditional form when time 
is orthogonal to space.

\section{SYNOPSIS OF NTO ANALYSIS}
\label{sec:synopsis}
\subsection{NTO Transformation and Metric}
As shown in references \cite{Langevin:1937}, \cite{Klauber:2003} and 
elsewhere, the rotating (NTO) frame metric can be derived using the most 
widely accepted transformation from the non-rotating (lab, upper case 
symbols) frame to a rotating (lower case) frame, i.e.,
\begin{equation}
\label{eq2}
\begin{array}{l}
 cT=ct \\ 
 R=r \\ 
 \Phi =\phi +\omega t \\ 
 Z=z \\ 
 \end{array}
\end{equation}
where $\omega $ is the angular velocity of the rotating frame and 
cylindrical spatial coordinates are used. The coordinate time $t$ for the 
rotating system equals the proper time of a standard clock located in the 
lab. The ultimate viability of this transformation rests on its capability 
to predict real world phenomena, as discussed at length in reference 
\cite{Klauber:2003}.

Substituting the differential form of (\ref{eq2}) into the line element in the lab 
frame
\begin{equation}
\label{eq3}
ds^2\mbox{ }=\mbox{ }-\mbox{ }c^2dT^2+\mbox{ }dR^2\mbox{ }+\mbox{ }R^2d\Phi 
^2\mbox{ }+\mbox{ }dZ^2
\end{equation}
results in the line element for the rotating frame
\begin{equation}
\label{eq4}
ds^2\;\;=\;-c^2(1-\textstyle{{r^2\omega ^2} \over 
{c^2}})dt^2\;+\;dr^2\;+\;r^2d\phi ^2\;+\;2r^2\omega d\phi 
dt\;+\;dz^2=\;g_{\alpha \beta } dx^\alpha dx^\beta .
\end{equation}
Note that the metric in (\ref{eq4}) is not diagonal, since $g_{\phi t} \ne 0$, and 
this implies that time is not orthogonal to space (i.e., an NTO frame.)

\subsection{Time in the Rotating Frame}
\label{subsec:mylabel1}
Time on a standard clock at a fixed 3D location on the rotating disk, found 
by taking \textit{ds}$^{2}= -c^{2}d\tau $ and \textit{dr = d}$\phi $ = \textit{dz = }0, is
\begin{equation}
\label{eq5}
d\tau =\sqrt {1-r^2\omega ^2/c^2} dt=\sqrt {-g_{tt} } dt,
\end{equation}
where \textit{dt} here indicates coordinate time passed at the same 3D location.

More generally, the coordinate time difference between two events at two 
infinitesimally separated locations on the disk (each having its own clock), 
in coordinate components, is \textit{dt}, where at least one of \textit{dr, d}$\phi $, \textit{dz} is not zero. 
The corresponding \textit{physical} time difference (i.e., time difference measured between 
two different standard clocks fixed in the rotating frame) in seconds can be 
found from \textit{dt} via the method described in Appendix C. It is
\begin{equation}
\label{eq6}
dt_{phys} =d\hat {t}=\sqrt {-g_{tt} } dt=\sqrt {1-r^2\omega ^2/c^2} dt,
\end{equation}
where the caret over $t$ is the common symbol for physical component. If the 
two locations in (\ref{eq6}) happen to be the same location, one obviously gets (\ref{eq5}).

Note that two events seen as simultaneous in the lab have \textit{dT} = 0 between them.  From the first line of (\ref{eq2}) and (\ref{eq6}), the same two events must also have 
$dt_{phys} =0$, and thus they are also seen as simultaneous on the disk. 
Thus, the lab and all disk standard clocks share common simultaneity. 
(Though standard disk clocks at different radii run at different rates and 
can not be synchronized, they can all agree that no time passed off any of 
them between two events.)

As discussed in references \cite{Klauber:2003} and \cite{Klauber:1998}, 
the author contends that there is only one possible simultaneity in a 
rotating frame that results in a single valued, continuous physical world 
time having all clocks in synchronization with themselves. That simultaneity 
is inherent in, and defined by, the transformation (\ref{eq2}).

\section{RADIAL AND CIRCUMFERENTIAL LIGHT SPEEDS}
\label{sec:radial}
In this section, we derive the speed of light (as measured with standard 
clocks and meter sticks) in a rotating frame according to NTO analysis.

\subsection{Circumferential Light Speed}
For light \textit{ds}$^{2}$ = 0. Inserting this into (\ref{eq4}), taking \textit{dr=dz=}0, and using the 
quadratic equation formula, one obtains a coordinate velocity (generalized 
coordinate spatial grid units per coordinate time unit) in the 
circumferential direction
\begin{equation}
\label{eq7}
v_{light,coord,circum} =\frac{d\phi }{dt}=-\omega \pm \frac{c}{r},
\end{equation}
where the sign before the last term depends on the circumferential direction 
(cw or ccw) of travel of the light ray.

The physical velocity (the value one would measure in experiment using 
standard meter sticks and clocks in units of meters per second) can be found 
from this via the method described in Appendix C. From that, one sees the 
local physical velocity is
\begin{equation}
\label{eq8}
v_{light,phys,circum} \;\;=\;\;\frac{d\hat {\phi }}{d\hat 
{t}}\;\;=\frac{\sqrt {g_{\phi \phi } } d\phi }{\sqrt {-g_{tt} } 
dt}\;\;=\;\;\frac{-\;r\omega \;\pm \;c}{\sqrt {1-\textstyle{{\omega ^2r^2} 
\over {c^2}}} }=\;\;\frac{-\;v\;\pm \;c}{\sqrt {1-\textstyle{{v^2} \over 
{c^2}}} },
\end{equation}
where $v$ is the circumferential speed of a point fixed in the rotating frame 
as seen from the lab. Note that for rotating frames, the local physical 
speed of light is not invariant or isotropic, and that this lack of 
invariance/isotropy depends on $\omega $, the angular velocity. Note 
particularly that this result is a direct consequence of the NTO nature of 
the metric in (\ref{eq4}). If $\omega $=0, physical (measured) light speed is 
isotropic and invariant, the metric is diagonal, and time is orthogonal to 
space.

If one accepts that the local, measured speed of light in rotating frames 
may be anisotropic, then (\ref{eq8}) is reasonable. As a first guess, one might 
consider the numerator as appropriate for such speed. Knowing that standard 
clocks on the rotating frame run more slowly than clocks in the lab, one 
would then consider modifying the first guess by the Lorentz factor, 
resulting in (\ref{eq8}).

If (\ref{eq8}) is correct, then the hypothesis of locality must be invalid for 
rotation, since a local Lorentz observer must measure a physical speed $c$ for 
light. As light speed is determined with meter sticks and standard clocks, 
the disk observer and the local co-moving Lorentz observer must see 
differences in their respective measurements for time and space. As 
discussed in references \cite{Klauber:2003} and \cite{Klauber:1998}, the 
author submits that the hypothesis of locality is only valid for reference 
frames in which time can be orthogonal to space without inducing 
discontinuities in time.

\subsection{Radial Direction Light Speed}
\label{subsec:radial}
For a radially directed ray of light, \textit{d$\phi $} = \textit{dz = }0, and \textit{ds} = 0 in (\ref{eq4}). Solving for 
\textit{dr/dt} one obtains
\begin{equation}
\label{eq9}
\frac{dr}{dt}=c\sqrt {1-r^2\omega ^2/c^2} .
\end{equation}
Since $g_{rr}$ = 1, the physical component, (measured with standard meter 
sticks) for radial displacement $d\hat {r}$ equals the coordinate radial 
displacement \textit{dr}. The local physical (measured) speed of light in the radial 
direction is therefore
\begin{equation}
\label{eq10}
v_{light,phys,radial} =\frac{d\hat {r}}{d\hat {t}}=\frac{\sqrt {g_{rr} } 
dr}{\sqrt {-g_{tt} } dt}=\frac{dr}{d\tau }=\frac{dr}{\sqrt {1-r^2\omega 
^2/c^2} dt}=c.
\end{equation}
\section{SAGNAC EXPERIMENT: THEORETICAL DERIVATION FROM ROTATING FRAME}
\label{sec:sagnac}
As Appendices A and B make clear, any analysis of the Sagnac effect from the 
point of view of the rotating frame (as opposed to the lab frame) must 
correctly predict different arrival times for the cw and ccw light pulses. 
This, of course, follows naturally in the NTO analysis, given that the cw 
and ccw pulses have different speeds. The derivation of the Sagnac effect, 
in terms of this time difference and the concomitant observed fringe shift, 
follows.

\subsection{Circular Light Paths about the Center of Rotation}
\label{subsec:circular}
The difference in time measured on the disk, between two pulses of light 
traveling opposite directions with respective physical speeds $v_+ $ and 
$v_- $ along a circumferential arc of length $l$, is
\begin{equation}
\label{eq11}
\Delta t_{phys} =\frac{l}{v_+ }-\frac{l}{v_- }
\end{equation}
Using (\ref{eq8}) this becomes
\begin{equation}
\label{eq12}
\begin{array}{c}
 \Delta t_{phys} =\frac{l\sqrt {1-v^2/c^2} }{c-v}-\frac{l\sqrt {1-v^2/c^2} 
}{c+v} \\ 
 \quad \\ 
 =\frac{2lv\sqrt {1-v^2/c^2} }{c^2-v^2}=\frac{v}{c}\frac{2l/c}{\sqrt 
{1-v^2/c^2} } \\ 
 \quad \\ 
 =\frac{v}{c}\frac{2t_{1way,c\;speed} }{\sqrt {1-v^2/c^2} } \\ 
 \end{array}.
\end{equation}
We will ignore the last line of (\ref{eq12}) in what follows, merely providing it 
here to show the relationship with the one way time for light over the 
distance $l$ for light speed = $c.$

To determine the time difference between two light pulses traveling once 
around the disk rim, we need to determine the length of the circumference as 
measured by an observer on the rotating disk. The physical measurement in 
meters for an infinitesimal distance along the circumference is
\begin{equation}
\label{eq13}
dl=d\hat {\phi }=\sqrt {g_{\phi \phi } } d\phi =rd\phi =\mbox{ 
distance}\;\mbox{measured}\;\mbox{in}\;\mbox{meter}\;\mbox{sticks}.
\end{equation}
Integrate (\ref{eq13}) from $\phi $ = 0 to 2$\pi $ to find the physical 
circumferential distance${\rm s}$ and one gets $l$ = 2$\pi r.$ Note this is \textit{not} 
equal to what one finds in the traditional analysis of rotating disks. In 
that analysis, circumferential Lorentz contraction is assumed to exist, but 
in NTO analysis, Lorentz contraction does not exist. That disparity is due 
ultimately to differences in simultaneity between the NTO and Lorentz 
transformations. (In traditional special relativity, Lorentz contraction is 
a result of differences in simultaneity between two translating frames.) For 
an extensive discussion on this topic, see Klauber\cite{Klauber:1998}.

Using (\ref{eq13}) in the second line of (\ref{eq12}), one obtains the time difference 
around the entire circumference between the cw and ccw pulses
\begin{equation}
\label{eq14}
\begin{array}{c}
 \Delta t_{phys} =\frac{v}{c}\frac{2l/c}{\sqrt {1-v^2/c^2} }=\frac{\omega 
r}{c^2}\frac{2(2\pi r)}{\sqrt {1-v^2/c^2} } \\ 
 \quad \\ 
 =\frac{4\omega A}{c^2\sqrt {1-v^2/c^2} } \\ 
 \end{array}.
\end{equation}
This equals the well known result derived from the lab frame shown in 
Appendix B. From it, one readily calculates the observed fringe shift for 
low circumferential velocity where \textit{v $<<$ c.} For that approximation, $c\Delta 
t_{phys} \approx \Delta \lambda $, and the variation in fringe location is
\begin{equation}
\label{eq15}
\Delta \lambda \approx c\Delta t_{phys} \approx \frac{4\omega A}{c}.
\end{equation}
(\ref{eq15}) can be readily generalized in terms of vector quantities 
\mbox{\boldmath$\omega$} and \textbf{\textit{A}}, and the fraction of fringe shift 
change $\Delta Z=\Delta \lambda /\lambda _0 $, where $\lambda _0 $ is the 
vacuum wavelength in the lab as
\begin{equation}
\label{eq16}
\Delta Z=\frac{\Delta \lambda }{\lambda _0 }\approx \frac{4\mbox{\boldmath$\omega$} \cdot 
\mbox{\boldmath$A$}}{c\lambda _0 },  
\end{equation}
which is the same as (\ref{eq1}).

\subsection{Arbitrary Closed Light Path}
\label{subsec:arbitrary}
An arbitrary closed path enclosing an area $A$ fixed on the rotating frame is 
depicted in Figure 1. For simplicity we first assume the entire area is 
planar and normal to the axis of rotation. 

\begin{figure}[htbp]
\centerline{\includegraphics[width=3.4in,height=1.96in]{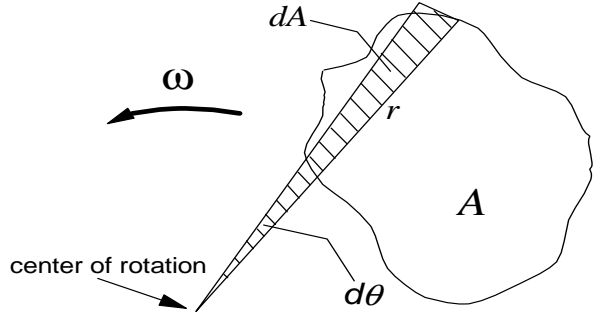}}
\caption{Arbitray Closed Area on Rotating Frame and the Sagnac Effect}
\label{fig1}
\end{figure}

From (\ref{eq12}) (or more precisely, from Appendix D), the first order difference 
in time between the cw and ccw light pulses over an infinitesimal length of 
the path is 
\begin{equation}
\label{eq17}
\begin{array}{c}
 d(\Delta t_{phys} )\approx \frac{v}{c^2}2(dl)=\frac{\omega r}{c^2}2(rd\phi 
)=\frac{2\omega \left( {2\textstyle{{r^2d\phi } \over 2}} \right)}{c^2} \\ 
 \quad \\ 
 =\frac{4\omega dA}{c^2}\;, \\ 
 \end{array}
\end{equation}
where \textit{dA} is the differential area (cross hatched in Figure 1) enclosed by the 
infinitesimal length of the path and the radial lines from the center of 
rotation to its endpoints. When one integrates this value over the entire 
path, the infinitesimal section over the same width infinitesimal angle on 
the opposite side of the area contributes a negative area to the 
integration. In effect, \textit{d$\theta $} has the same absolute value, but opposite sign. 
Upon completing the integration the net area left is $A,$ that enclosed by the 
closed path. 

The derivation readily generalizes to cases where the axis of rotation is 
not normal to the area enclosed and results in
\begin{equation}
\label{eq18}
\Delta t_{phys} \approx \frac{4\mbox{\boldmath$\omega$} \cdot \mbox{\boldmath$A$}}{c^2}.
\end{equation}
Hence, (\ref{eq16}) is a very general relation, accurate to first order, and valid 
for any enclosed area, as was determined by experiment and noted by 
Post\cite{Post:1967}.

\section{SUMMARY AND CONCLUSION}
NTO analysis differs from traditional approaches by contending that the 
continuous and single valued nature of physical time constrains simultaneity 
in a rotating frame to be unique (and not a matter of convention.) It 
thus predicts anisotropic, local, physical light speed in rotating frames, 
and correctly predicts the Sagnac and related thought experiment results 
from the point of view of the rotating frame observer. Time arrival 
differences found between cw and ccw light pulses, as well as the associated 
experimentally verified fringe shifting, are shown to precisely match the 
values determined from lab frame based analysis. Inherent in NTO analysis is 
the invalidity of the hypothesis of locality for rotation, as well as the 
prediction that there is no circumferential Lorentz contraction effect on a 
rotating disk. In these matters, NTO analysis differs from the traditional 
analysis of rotating systems.

\section{ACKNOWLEDGEMENTS}
The author thanks three referees for valuable suggestions. Their comments 
led to a manuscript that is substantially clearer, and more relevant by 
contrast with the traditional approach to the Sagnac issue, than the 
original submission.

\section*{APPENDIX A -- THOUGHT EXPERIMENT AND SAGNAC}
The speed of light, according to the first relativity postulate, must be 
measured the same by any observer, under any conditions. Keeping the first 
postulate in mind, consider the following thought experiment involving an 
observer fixed to the rotating disk of Figure 2 who measures the speed of 
light.

The observer shown has already laid meter sticks along the rim circumference 
and determined the distance around that circumference. As part of his 
experiment, he has also set up a cylindrical mirror, reflecting side facing 
inward, all around the circumference. He takes a clock with him and anchors 
himself to one spot on the disk rim. When his clock reads time $T$ = 0 (left 
side of Figure 2) he shines two short pulses of light (the mini sine waves 
in the figure with accompanying arrows indicating direction) tangent to the 
rim in opposite directions. The mirror will cause these light pulses to 
travel circular paths around the rim, one clockwise (cw) and one 
counterclockwise (ccw).

\begin{figure}[htbp]
\centerline{\includegraphics[width=4.1in,height=1.59in]{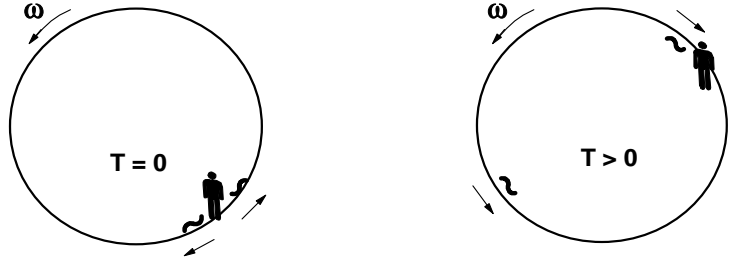}}
\caption{Rotating Disk Observer Measuring Light Speed}
\label{fig2}
\end{figure}

From the ground we see the cw and ccw light pulses having the same speed 
$c$, the usual value for the speed of light. Note, however, that as the pulses 
travel around the rim, the rim and the observer fixed to it move as well. 
Hence, a short time later, as illustrated in the right side of Figure 2, the 
cw light pulse has returned to the observer, whereas the ccw pulse has yet 
to do so. A little later (not shown) the ccw pulse will have caught up to 
the observer.

For the observer, from his perspective on the disk, both light rays travel 
the same distance, the same number of meters around the circumference. But 
his experience and his clock readings tell him that the cw pulse took less 
time to travel the same distance around the circumference than the ccw 
pulse.

What can he conclude? It appears he can only conclude that, from his point 
of view, the cw pulse traveled faster than the ccw pulse. Hence, it seems 
the speed of light as measured on the rotating disk does not always have the 
same value. It appears different in different directions, and different from 
that measured on the ground.

This thought experiment makes it plain that any explanation for the Sagnac 
experiment, from the point of view of the disk reference frame, must account 
for different \textit{arrival times} for the cw and ccw light pulses. Analyses based on Doppler 
shifts or wave length changes are simply not sufficient to explain this. 
This conclusion accords with 
GPS\cite{Ashby:2002}$^{,}$\cite{Ashby:1997} and other 
data\cite{Saburi:1976} for the rotating frame of the earth.

\section*{APPENDIX B -- DERIVING SAGNAC RESULT FROM THE LAB FRAME}
\label{sec:appendix}
Consider Figure 2 of Appendix A with time ($T>0)$ in the right side of the 
figure when the cw light pulse reaches the disk observer designated as 
$T_{1}$. Label the time when the ccw pulse reaches the disk observer (not 
shown) as $T_{2}$. Then lengths traveled as seen from the lab by the ccw 
light pulse and the observer at $T_{1}$ must sum to equal the circumference, 
i.e.
\begin{equation}
\label{eq19}
cT_1 +\omega RT_1 =2\pi R\quad \quad \to \quad \quad T_1 =\frac{2\pi 
R}{c+\omega R}.
\end{equation}
Similarly, at time $T_{2}$ 
\begin{equation}
\label{eq20}
cT_2 =\omega RT_2 +2\pi R\quad \quad \to \quad \quad T_2 =\frac{2\pi 
R}{c-\omega R}.
\end{equation}
Hence,
\begin{equation}
\label{eq21}
\Delta T=T_2 -T_1 =\frac{2\omega R}{c^2}\frac{2\pi R}{(1-\omega 
^2R^2/c^2)}\approx \frac{4\omega A}{c^2}.
\end{equation}
As is well known, the standard (physical) clocks on the disk rim run more 
slowly than the lab clocks by $\sqrt {1-\omega ^2R^2/c^2} $, so the observer 
on the disk must measure an arrival time difference of
\begin{equation}
\label{eq22}
\Delta t_{phys} =\frac{4\omega A}{c^2\sqrt {1-\omega ^2R^2/c^2} }.
\end{equation}
Steps identical to (\ref{eq15}) and (\ref{eq16}) yield (\ref{eq1}).

\section*{APPENDIX C. PHYSICAL VS COORDINATE COMPONENTS}
\label{sec:mylabel2}
If one has coordinate components, found from generalized coordinate tensor 
analysis, for some quantity, such as stress or velocity, one needs to be 
able to translate those into the values measured in experiment. For some 
inexplicable reason, the method for doing this is not typically taught in 
general relativity (GR) texts/classes, so it is reviewed here. (Note that 
often in GR, one seeks invariants like $d\tau $, \textit{ds}, etc., which are the same 
in any coordinate system, and in such cases, this issue does not arise. The 
issue does arise with vectors/tensors, whose coordinate components vary from 
coordinate system to coordinate system.)

The measured value for a given vector component, unlike the coordinate 
component, is unique within a given reference frame. In differential 
geometry (tensor analysis), that measured value is called the ``physical 
component''.

Many tensor analysis texts show how to find physical components from 
coordinate components\cite{The:1972}. A number of continuum mechanics texts 
do as well\cite{Malvern:1988}. The only GR text known to the present author 
that mentions physical components is Misner, Thorne, and 
Wheeler\cite{Misner:1973}. Those authors use the procedure to be described 
below, but do not derive it\cite{Physical:1}. The present author has 
written an introductory article on this, oriented for students, that may be 
found at the Los Alamos web site\cite{Klauber:6}. The following is 
excerpted in part from that article.

The displacement vector $d$\textbf{x }between two points in a 2D Cartesian 
coordinate system is
\begin{equation}
\label{eq23}
d{\rm {\bf x}}\,=dX^1{\rm {\bf \hat {e}}}_1 +dX^2{\rm {\bf \hat {e}}}_2 
\end{equation}
where the ${\rm {\bf \hat {e}}}_i $ are unit basis vectors and \textit{dX}$^{i}$ are 
physical components (i.e., the values one would measure with meter sticks). 
For the same vector $d$\textbf{x} expressed in a different, generalized, 
coordinate system we have different coordinate components \textit{dx}$^{i }\ne $ 
\textit{dX}$^{ i }$ (\textit{dx}$^{i }$do not represent values measured with meter sticks), but a 
similar expression
\begin{equation}
\label{eq24}
d{\rm {\bf x}}\,=dx^1{\rm {\bf e}}_1 +dx^2{\rm {\bf e}}_2 ,
\end{equation}
where the generalized basis vectors \textbf{e}$_{i }$point in the same 
directions as the corresponding unit basis vectors ${\rm {\bf \hat {e}}}_i 
$, but are not equal to them. Hence, for ${\rm {\bf \hat {e}}}_i $, we have
\begin{equation}
\label{eq25}
{\rm {\bf \hat {e}}}_i =\frac{{\rm {\bf e}}_i }{\vert {\rm {\bf e}}_i \vert 
}=\frac{{\rm {\bf e}}_i }{\sqrt {{\rm {\bf e}}_{\underline{i}} \cdot {\rm 
{\bf e}}_{\underline{i}} } }=\frac{{\rm {\bf e}}_i }{\sqrt 
{g_{\underline{i}\underline{i}} } }
\end{equation}
where underlining implies no summation.

Substituting (\ref{eq25}) into (\ref{eq23}) and equating with (\ref{eq24}), one obtains
\begin{equation}
\label{eq26}
dX^1=\sqrt {g_{11} } dx^1\quad \quad \quad dX^2=\sqrt {g_{22} } dx^2,
\end{equation}
which is the relationship between displacement physical (measured with 
instruments) and coordinate (mathematical value only) components.

Consider a more general case of an arbitrary vector \textbf{v} 
\begin{equation}
\label{eq27}
{\rm {\bf v}}=v^1{\rm {\bf e}}_1 +v^2{\rm {\bf e}}_2 =v^{\hat {1}}{\rm {\bf 
\hat {e}}}_1 +v^{\hat {2}}{\rm {\bf \hat {e}}}_2 
\end{equation}
where, \textbf{e}$_{1}$ and \textbf{e}$_{2}$ here do not, in general, have 
to be orthogonal, \textbf{e}$_{i}$ and ${\rm {\bf \hat {e}}}_i $ point in 
the same direction for each index $i$, and carets over component indices 
indicate physical components. Substituting (\ref{eq25}) into (\ref{eq27}), one readily 
obtains
\begin{equation}
\label{eq28}
v^{\hat {i}}=\sqrt {g_{\underline{i}\underline{i}} } v^i,
\end{equation}
which we have shown here to be \textit{valid in both orthogonal and non-orthogonal systems}.

As a further aid to those readers familiar with anholonomic coordinates 
(which are associated with non-coordinate basis vectors superimposed on a 
generalized coordinate grid), we note that physical components are special 
case anholonomic components where the non-coordinate basis vectors have unit 
length.

It is important to note that anholonomic components do not transform as true 
vector components. So one can not simply use physical components in tensor 
analysis as if they were. Typically, one starts with physical components as 
input to a problem. These are converted to coordinate components, and the 
appropriate tensor analysis carried out to get an answer in terms of 
coordinate components. One then converts these coordinate components into 
physical components as a last step, in order to compare with values measured 
with instruments in the real world.

As a basis vector is derived from infinitesimals (derivative at a point), 
one sees (\ref{eq28}) is valid locally in curved, as well as flat, spaces, and can 
be extrapolated to 4D general relativistic applications. So, very generally, 
for a 4D vector $v^{\mu }$
\begin{equation}
\label{eq29}
v^{\hat {i}}=\sqrt {g_{\underline{i}\underline{i}} } v^i\quad \quad v^{\hat 
{0}}=\sqrt {-g_{00} } v^0,
\end{equation}
where Roman sub and superscripts refer solely to spatial components (i.e. 
$i$ = 1,2,3.)

\section*{APPENDIX D -- GENERAL PATH DIRECTIONAL TIME DIFFERENCE FOR LIGHT}
\label{sec:mylabel3}
Relation (\ref{eq8}) for the circumferential physical velocity of light on a 
rotating frame in terms of the circumferential non-rotating frame (lab) 
velocity generalizes to any such velocity as\cite{Ref:1}
\begin{equation}
\label{eq30}
v_{circum,phys,rot} \,\;=\frac{-r\omega \pm v_{circum,phys,lab} }{\sqrt 
{1-r^2\omega ^2/c^2} }.
\end{equation}
We need to find the velocity of light along the path length \textit{dl} in Figure 3 
where the path followed by light is not along a circumferential arc having 
its center at the axis of rotation. To this end consider \textit{dl} aligned at an 
angle $\alpha $ to such a circumferential arc as shown in Figure 3. In order 
to simplify otherwise unwieldy algebra we will restrict ourselves to first 
order relations from the outset.

The light travels along \textit{dl} so we need to find the speed of light along this 
path in the rotating frame in the positive and negative directions. As 
before, we label such speeds as $v_{+}$ and $v_{-}$ . According to our first 
order assumption, we further consider the angle $\alpha $ to have the same 
value in the lab and rotating frames.

\begin{figure}[htbp]
\centerline{\includegraphics[width=2.46in,height=1.93in]{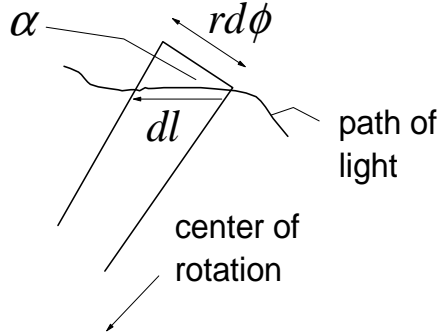}}
\caption{Path of Light Pulses and a Circumferential Element}
\label{fig3}
\end{figure}

With the aid of (\ref{eq30}), and the understanding implicit in (\ref{eq10}) that the 
velocity of light in the radial direction is the same in the rotating frame 
as in the lab, we find
\begin{equation}
\label{eq31}
\begin{array}{l}
 v_{light,circum,lab} =c\cos \alpha \quad \quad \quad \quad \quad 
v_{light,radial,lab} =c\sin \alpha \\ 
 \quad \\ 
 v_{light,circum,rot} \approx -v\pm c\cos \alpha \quad \quad \quad \quad \quad
v_{light,radial,rot} =c\sin \alpha \\ 
 \quad \\ 
 v_\pm =v_{light,along\;dl,rot} =\sqrt {(v_{light,circum,rot} 
)^2+(v_{light,radial,rot} )^2} \approx \sqrt {v^2+c^2\mp 2cv\cos \alpha } \\ 
 \quad \\ 
 \end{array}
\end{equation}
Proceeding in parallel fashion to (\ref{eq11}) to (\ref{eq14}), we find
\begin{equation}
\label{eq32}
\begin{array}{c}
 d(\Delta t_{phys} )=\frac{dl}{v_+ }-\frac{dl}{v_- }\approx \frac{dl}{\sqrt 
{v^2+c^2-2cv\cos \alpha } }-\frac{dl}{\sqrt {v^2+c^2+2cv\cos \alpha } } \\ 
 \quad \\ 
 \approx \frac{dl}{c\sqrt {1-2\textstyle{v \over c}\cos \alpha } 
}-\frac{dl}{c\sqrt {1+2\textstyle{v \over c}\cos \alpha } } \\ 
 \quad \\ 
 \approx \frac{dl}{c}(1+\textstyle{v \over c}\cos \alpha -1+\textstyle{v 
\over c}\cos \alpha ) \\ 
 \quad \\ 
 =\frac{v}{c^2}2dl\cos \alpha \\ 
 \end{array}
\end{equation}
We also know
\begin{equation}
\label{eq33}
dl=\frac{rd\phi }{\cos \alpha }.
\end{equation}
Hence
\begin{equation}
\label{eq34}
d(\Delta t_{phys} )\approx \frac{v}{c^2}2rd\phi =\frac{2\omega r^2d\phi 
}{c^2},
\end{equation}
our starting point in Section \ref{subsec:arbitrary}.

\end{document}